\documentclass{article}
\usepackage{stywhispers,amsmath,epsfig}
\usepackage{amsmath,amsfonts,amssymb}
\usepackage{graphicx,makecell}
\usepackage[colorlinks=true, allcolors=blue]{hyperref}
\usepackage{subcaption}
\usepackage{tabularray}
\usepackage{float}
\usepackage{graphicx}
\usepackage{float}
\usepackage{amsmath}

\title{Theoretical and Practical Progress in Hyperspectral Pixel Unmixing with Large Spectral Libraries from a Sparse Perspective}
\twoauthors
  {Jade Preston}
	{University of Virginia\\
	School of Data Science\\
	Charlottesville, VA}
  {William Basener}
	{University of Virginia\\
	School of Data Science\\
	Charlottesville, VA}
\begin{document}
%
\maketitle
\begin{abstract}
Hyperspectral unmixing is the process of determining the presence of individual materials and their respective abundances from an observed pixel spectrum. Unmixing is a fundamental process in hyperspectral image analysis, and is growing in importance as increasingly large spectral libraries are created and used. Unmixing is typically done with ordinary least squares (OLS) regression. However, unmixing with large spectral libraries where the materials present in a pixel are not a priori known, solving for the coefficients in OLS requires inverting a non-invertible matrix from a large spectral library. A number of regression methods are available that can produce a numerical solution using regularization, but with considerably varied effectiveness.  Also, simple methods that are unpopular in the statistics literature (i.e. step-wise regression) are used with some level of effectiveness in hyperspectral analysis. In this paper, we provide a thorough performance evaluation of the methods considered, evaluating methods based on how often they select the correct materials in the models. Investigated methods include ordinary least squares regression, non-negative least squares regression, ridge regression, lasso regression, step-wise regression and Bayesian model averaging. We evaluated these unmixing approaches using multiple criteria: incorporation of non-negative abundances, model size, accurate mineral detection and root mean squared error (RMSE). We provide a taxonomy of the regression methods, showing that most methods can be understood as Bayesian methods with specific priors.  We conclude that methods that can be derived with priors that correspond to the phenomenology of hyperspectral imagery outperform those with priors that are optimal for prediction performance under the assumptions of ordinary least squares linear regression.
\end{abstract}


\section{INTRODUCTION}
\label{intro} 
Hyperspectral Images are images with many, often hundreds, of contiguous wavelengths of light. Thus, each pixel contains a spectrum for the materials present at the location imaged. Hyperspectral images have been crucial in numerous fields, including but not limited to mineral detection, agricultural and environmental monitoring, pollution surveillance, medicine, and water purity assessment \cite{bioucas2013hyperspectral,borsoi2021spectral,rasti2023image}.                                                                     
The ground sample distance (GSD) in an image is the distance between the ground locations corresponding to the centers of neighboring pixels.  If we imagine the region on the ground corresponding to a pixel as a rectangle, the GSD corresponds to the length of a side of such a rectangle and is often thought of as a measure of pixel size or an inverse measure of spatial resolution. (The region on the ground corresponding to a pixel is more accurately modeled as an ellipse or 2-dimensional Gaussian.) Because the light entering a hyperspectral sensor is separated into many wavelengths, the GSD of a hyperspectral sensor is higher than the GSD for a multispectral sensor of comparable construction and noise levels, which is yet higher GSD than a comparable panchromatic imager. Because the GSD of hyperspectral sensors are generally large, the spectrum measured in a pixel is often a mixture of the spectra for individual pure materials present.

\subsection{Linear and Nonlinear Mixture Models}
If only one material is present on the ground at the location measured by a pixel, the pixel is called a ``pure pixel'' and the spectrum for the pixel will be the spectrum for the given material.  However, the large pixel size of hyperspectral imagery and phenomenology involved imply that in most situations pure pixels are rare, and the measured spectrum will be a mixture of the spectra of the individual materials present.  This is the case, for example, where the region on the ground measured in the pixel covers both a car and asphalt road surface around the car. In the idealized model for this case the observed pixel spectrum would be the linear mixture $\mathbf{s} = a_c \mathbf{s_c} + a_r \mathbf{s_r}$ where $\mathbf{s_c}$ and $\mathbf{s_r}$ are the spectra of the car and road, and $a_c$ and $a_r$ are the fraction of the area measured in the pixel occupied by the car and road, respectively.  The values $a_c$ and $a_r$ are called the fractional abundances~\cite{nash1974spectral} for the respective materials.

Our car on a road is an example of the linear mixture model, which is given by the formula
\begin{equation}
\label{eqn:linearMixingModel}
\mathbf{y} = \sum_{i} a_i \mathbf{s}_i + \boldsymbol{\varepsilon}.
\end{equation}
In this equation, $\mathbf{y}$ is a pixel spectrum, the $\mathbf{s}_i$ are spectra for the materials present at the location of the pixel, the $a_i$ are their relative fractional abundances, and $\mathbf{\varepsilon}$ is a noise error term. The process of unmixing using this model is a process of determining spectra $\mathbf{s}_i$ and coefficients $a_i$ that result in small error. If the materials that are present in the pixel are known, then unmixing only requires finding the abundances that minimize the error in Equation~\ref{eqn:linearMixingModel}.  In this case, ordinary least squares regression provides the formula for the coefficients. However, usually the set of materials in a pixel are not known prior to analysis. Then the $\mathbf{s}_i$ must be selected from an often large library of potential materials. Ordinary least squares regression is not generally useful for determining which spectra to select for the model.

The spectra $\mathbf{s}_i$ are sometimes called endmembers, particularly in the context of linear unmixing of every pixel in an image with a single fixed set of $\mathbf{s}_i$. If $N$ is the number of endmembers, then the scatterplot of the image spectra will form an $N-1$-dimensional simplex with the $\mathbf{s}_i$ as the vertices of the simplex. In early research on unmixing, the $\mathbf{s}_i$ spectra were determined from the image~\cite{keshava2003survey} by locating the vertices of a simplex modeling the image assuming a linear mixture model. This provides a model for the image similar to decomposing the image into eigenvectors capturing most of the information. Typically a small number of endmembers are desirable, say 5-10, to provide a concise model for the image.  We will use the term endmember only for the spectra $\mathbf{S}_i$ which are pixel spectra from an image. Since our emphasis is on unmixing with spectra of known materials from a library, $\mathbf{S}_i$, we will generally refer to the $\mathbf{S}_i$ as constituent spectra.

There are situations where the function for approximating $\mathbf{y}$ from the constituents $\mathbf{s}_i$ (the right-hand side of Equation~\ref{eqn:linearMixingModel}) is not linear.  The process of determining the coefficients or parameters for the function is called (nonlinear) unmixing. The most common form of nonlinear unmixing is using a polynomial function, in which the higher order terms model multi-bounce photon trajectories. The coefficients of the higher order correspond to the prevalence of photon interactions.  While the polynomial regression formula is nonlinear with respect to the constituent spectra $\mathbf{s}_i$, it is linear with respect to the individual terms, and thus readily solvable when the number of distinct constituent spectra is small.

As spectral libraries have become more common, unmixing with spectra from a library rather than image pixels has become substantially more important in practical applications.  Unmixing with library spectra provides precise information that may be essential for the user, for example the specific mix of minerals present for a geological understanding of an area, or a specific chemical pollutant that may be present in sand or soil. As such, unmixing with a library provides a process for subpixel material identification~\cite{loughlin2020efficient}.

However, this increase in potential information increase comes with a cost.  Solving for the coefficients in the the linear mixing model when there are more spectra in the library than bands in the image using the ordinary least squares formula requires inverting a non-invertible matrix. Thus, unmixing with large libraries requires some form of reducing the number of materials that will be in the final model. Methods for regression that provide a limited number of terms in the final model are sometimes called sparse regression, regression with regularization, or model selection. 

The increase in library size and complexity is not simply an academic problem.  For example, soil was historically often considered a pure endmember category in the image unmixing paradigm, but soil is in fact very complex~\cite{mitchell2005fundamentals, richter2012changing}.  As of 2016, the ``global vis–NIR soil spectral library''~\cite{rossel2016global} contained over 23,632 distinct spectra measured across 350nm-2500nm from 92 different countries, and almost all spectra contain substantial metadata. The scale, variety, and specificity of libraries like this are outpacing exploitation methods and our understanding of how to utilize these methods on such large and detailed VNIRSWIR spectroscopy data.  

It is the authors' hope that -- in addition to providing theoretically and experimentally supported recommendations for unmixing currently -- the approach to investigating unmixing methods in this paper provide a sound theoretical framework and experimental support for advancing exploitation methods generally, including unmixing and identification with large complex information-rich libraries.

\subsection{Phenomenology of mixing}

\begin{figure}[ht] 
\begin{center}
   \begin{subfigure}{.39\textwidth}
    \includegraphics[width=\textwidth]{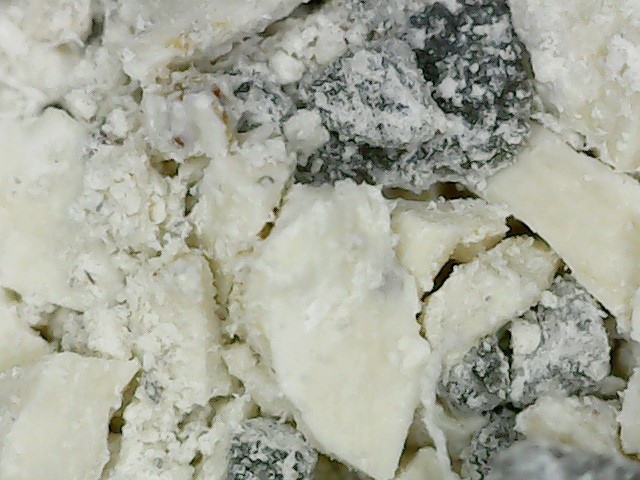}
    \caption{}
    \label{fig:sub1}
    \end{subfigure}
    \begin{subfigure}{.39\textwidth}
    \includegraphics[width=\textwidth]{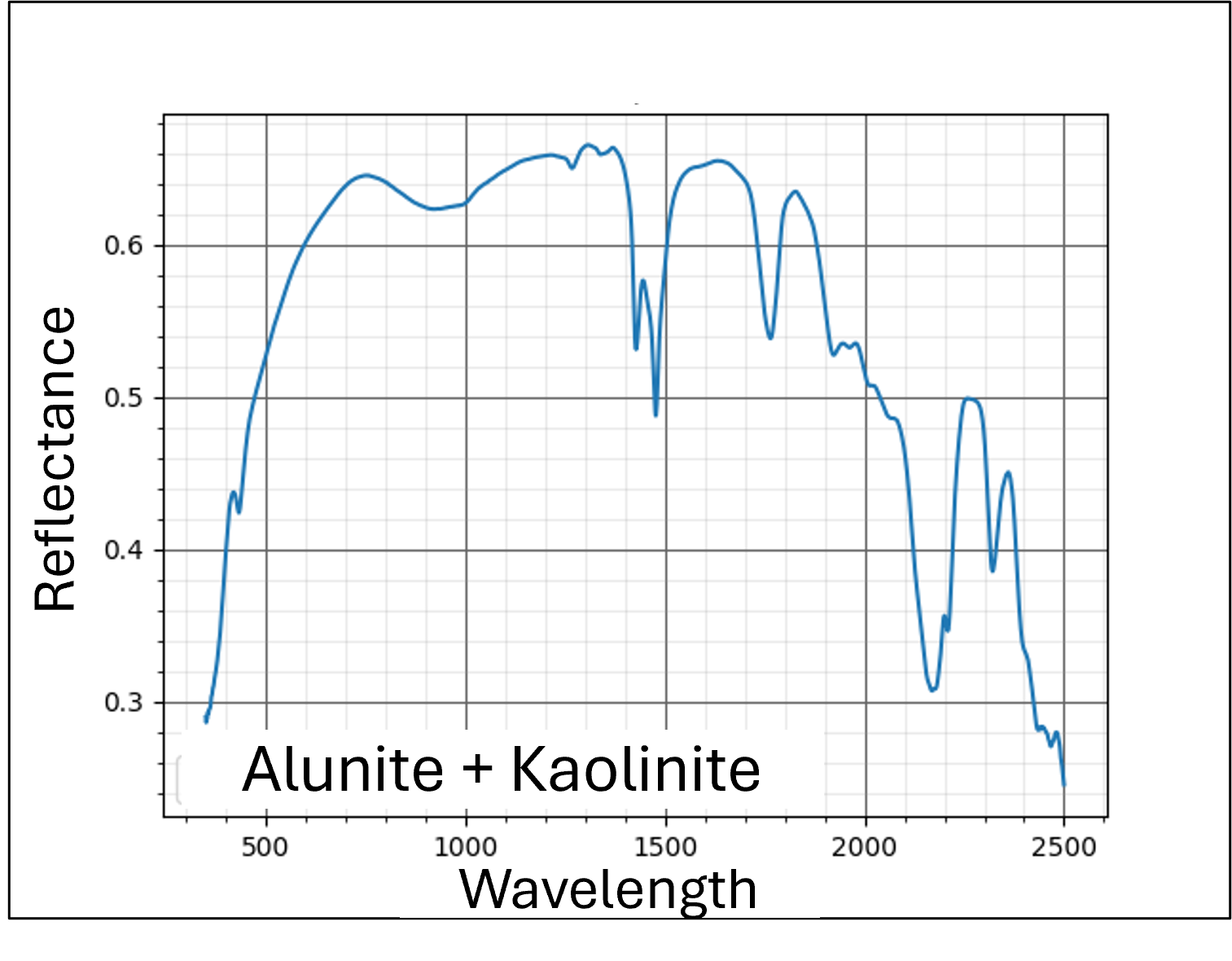}
    \caption{}
    \label{fig:sub2}
    \end{subfigure} 
\caption{\label{fig:1} Figure \ref{fig:sub1} is an example of a mineral mixture of alunite and kaolinite. Figure \ref{fig:sub2} is a spectrum from the mixture obtained using an ASD field spectrometer.
}
\end{center}
\end{figure}

Figure \ref{fig:1} provides an example of an intimate mixture of minerals (Figure \ref{fig:sub1}) and an associated mixed pixel spectrum (Figure \ref{fig:sub2}). 

Observed mixed pixel spectra can be described by their endmembers and associated fractional abundances . The process of identifying the pure materials in a pixel -- using spectra mapped over hundreds or thousands of wavebands -- and their estimating material abundances is called hyperspectral unmixing \cite{keshava2002spectral, wei2020overview}. Delineation between endmembers can be somewhat subjective \cite{bioucas2012hyperspectral} and dependent on the goal of unmixing. For example, if we have a pixel containing both soil and vegetation, we could say there are two endmembers in the scene. Alternatively, we could increase our enumeration to include the various minerals within the soil as well as the multiple types of vegetation also within the pixel \cite{keshava2002spectral, bioucas2012hyperspectral}. Often by increasing the number of endmembers in a pixel, the unmixing process can grow in complexity.

Researchers categorize hyperspectral unmixing algorithms into two approaches: linear mixture modeling and nonlinear mixture modeling \cite{ibarrola2019hyperspectral}. Modeling-approach determination depends on the elements in the scene and the goal for unmixing \cite{wei2020overview}. Though possible (and at times more efficient), there have been fewer attempts to unmix intimate mixtures -- i.e. mineral deposits -- using linear unmixing methods. The primary method for unmixing intimately mixed components is nonlinear mixture models. With complex mixtures, endmembers are often not known a priori and require large spectral libraries for variable comparison \cite{borsoi2021spectral}. Shading in the region, atmospheric effects and or differences in material structure can cause spectral variability between pixels. Therefore, regularization in the modeling process is a necessity. Linear mixture models can provide certain advantages such as faster computation and directly interpretable results \cite{keshava2002spectral,keshava2000algorithm}. Moreover, many linear mixture model techniques do not require the use of dimensionality reduction methods because of their inherent regularization in the form of penalties on model size \cite{li2019local}.  

Few research papers compare partially constrained least squares, sparse regression, iterative feature selection, Bayesian model averaging (BMA) \& nonlinear methods \cite{gault2016comparing}. Those that do make this comparison, consider a subset of algorithms \& lack a holistic view of their trade-offs. Additionally, no studies provide this comparison in terms of physical-chemical feature detection of material types.

This study compares an assortment of effective unmixing approaches for mineral unmixing which incorporate regularization components, and realistic modeling constraints. The approaches compared in this study include ordinary least squares regression, non-negative least squares (NNLS) regression, ridge regression, lasso regression, step-wise regression and BMA. Additionally, this paper will contribute a comprehensive overview and discussion regarding technique performance involved in positively identifying the minerals.

In the preceding sections of this research paper, we examine essential aspects for advancing hyperspectral unmixing of intimate mixtures. Section \ref{methods} describes the dataset derivation  and details the techniques employed in this study. The results from this methodology using our derived dataset are presented in Section \ref{results}. Section \ref{discussion} provides a detailed discussion of the results and implications.

\begin{figure*}[h!]
    \centering
    \includegraphics[width=.9\linewidth]{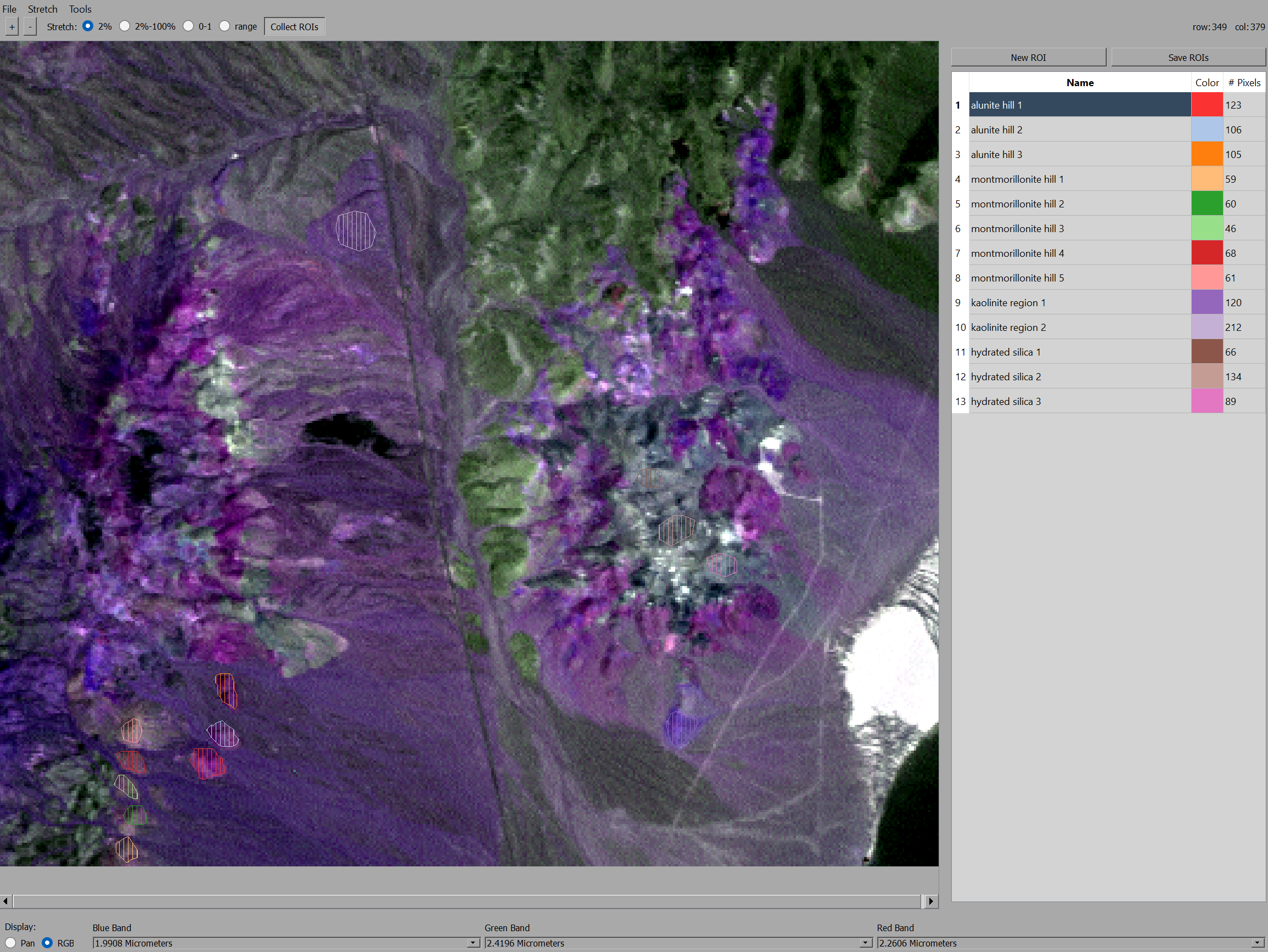}
    \caption{\label{hyperspectralpy} This figure displays the selected pixels using ``Hyperspectralpy" \cite{Basener_2022b}. The selected regions are labeled based on the primary mineral of interest for unmixing in that area.}
\end{figure*}

\begin{table*}[h!] 
\begin{center}    \includegraphics[width=.9\textwidth]{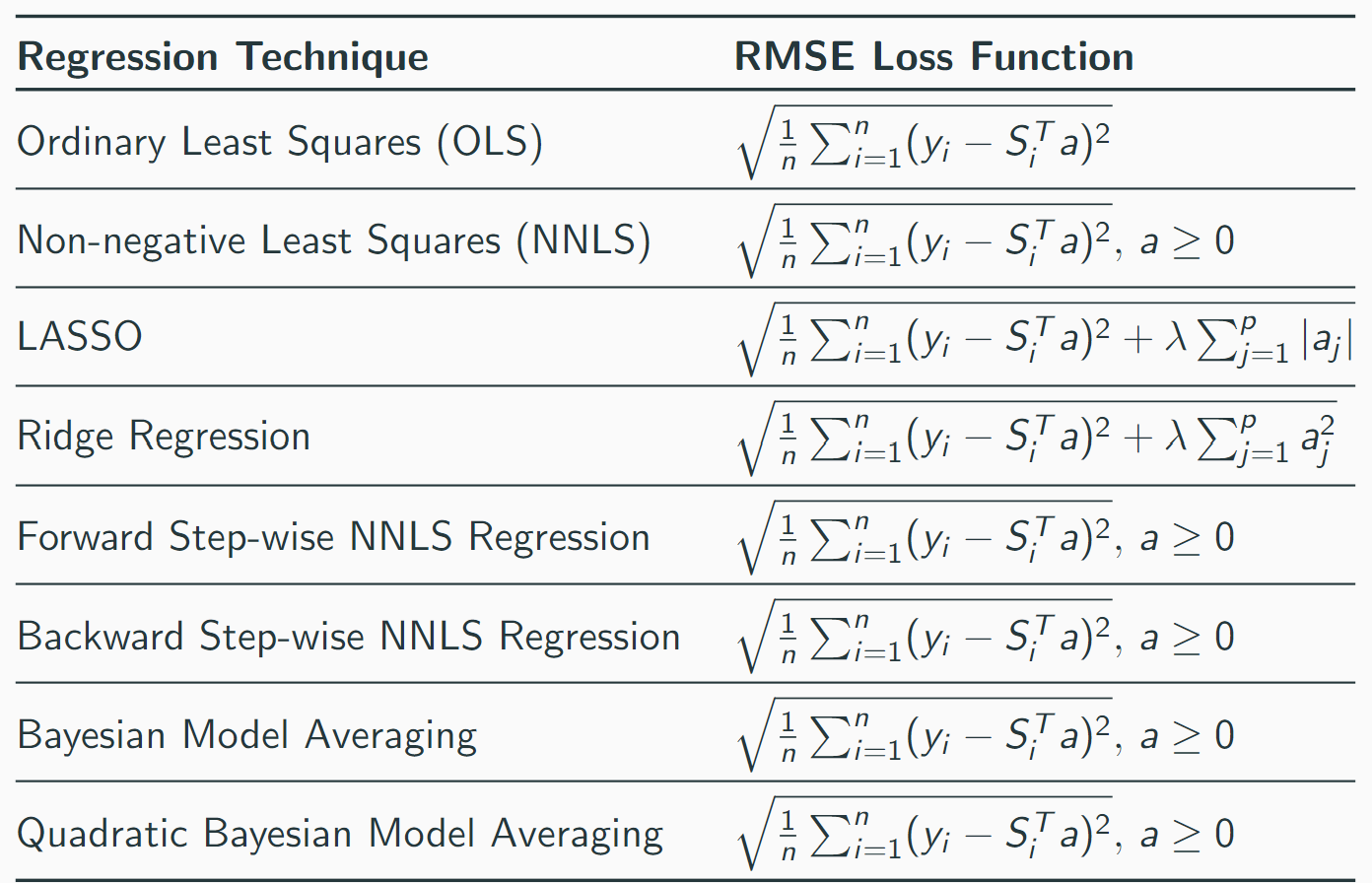}
\caption{\label{lossfunction} This figure displays the loss function for each technique incorporated.
}
\end{center}
\end{table*}

\section{Methods}
\label{methods}
\subsection{Dataset}
We use the USGS splib07a spectral library which has 481 spectra of known minerals~\cite{clark2007usgs,swayze2014mapping} and collected additional spectra with an ASD Field Spectrometer. We tested unmixing methods on a hyperspectral image collected with the NASA Jet Propulsion Laboratory Airborne Visible/Infrared Imaging Spectrometer (AVIRIS) over the well studied Cuprite hills mineralogical site. AVIRIS collects spectra in 224 continuous bands mapping over the wavelength range 0.38-2.5 micrometers (\(\mu m\)). We focus on a subset of the AVIRIS image consisting of 400 x 350 pixels, and used 50 bands with wavelengths ranging 2 to 2.5 micrometers.

We collected the observed pixels -- regions of interest (ROIs) -- using the python package ``Hyperspectralpy". This package exports the selected pixels' location (in the image) and associated spectral information for selected pixels to a csv file \cite{Basener_2022b}. Figure \ref{hyperspectralpy} shows the ROIs within the dataset. The legend displays the contents of the observation data. Although there are more selected regions, the alunite and kaolinite ROIs were used for this unmixing study.

To evaluate unmixing performance for each technique, model error was computed using RMSE we selected two pixel regions with the goal of successfully detecting the primary mineral associated to that pixel location (alunite and kaolinite). Successful detection equated to reasonable abundance quantification (\(0.1 \leq a_i\)) assigned to the mineral's correlated spectra. The target mineral was considered undetected if its assigned abundance was less than 0.1.

\subsection{Least Squares Regression}

Unconstrained OLS Regression serves as a baseline for many unmixing studies, but it is known for providing impractical or suboptimal results \cite{heinz2021supervised}. We incorporate the OLS unmixing results in this study to present a holistic glimpse surrounding the progression of this research. The unconstrained OLS model means our model had no requirements associated with constraining the material abundances. Typically, there are two constraints implemented on the linear model: abundance-sum-to-one constraint and non-negativity constraint \cite{heinz2001fully,heinz2021supervised}. Material abundances in this model could assume negative values and a sum greater than one. We utilized the sklearn ``LinearRegression" package to implement the unconstrained OLS unmixing \cite{Gramfort_2024}. 

NNLS is an extention of ordinary least squares regression. In this study, the NNLS technique solely constrains the abundances to be non-negative. We avoided constraining the abundances to be less than 1. To incorporate this NNLS technique, we implemented the ``NNLS" python package from scipy.optimize \cite{nnls-SciPyv1.14.0Manual}.

\subsection{Regularization Techniques}
Sparse unmixing becomes necessary when the spectral library is a large dataset comprised of a combination of scene endmembers and the observed pixel is a mixture of a far smaller subset of materials from that library \cite{iordache2011sparse}. Researchers have explored sparse regression models such as lasso regression, ridge regression or a combination of the two. Ridge regression and lasso regression are extensions of ordinary least squares with the L2 and L1 penalty norms weighting the material abundances respectively \cite{li2019local,iordache2013collaborative}. The L2 norm is also referred to euclidean distancing while the L1 norm is manhattan distancing. Additionally, it is important to note that these regularization techniques are related to Bayesian regression. Ridge regression is bayes optimal when the parameters follow a Gaussian prior distribution. Lasso regression is bayes optimal when the parameters follow a Laplace prior distribution \cite{bedoui2020bayesian}. We implemented ridge regression and lasso regression individually to uncover any unmixing strengths related to the individual approaches. ``Ridge" and ``Lasso" python packages from sklearn were used for implementation \cite{scikit_ridge2024,scikit_lasso2024}.

\subsection{Interactive Approaches}
Researchers have also achieved unmixing success through interactive approaches such as step-wise regression \cite{gault2016comparing,winter2003examining}. We implemented both forward (FSR) and backward (BSR) step-wise regression to ascertain the unmixing advantages each model provides. The code for FSR and BSR were developed independently rather than a python package. Both techniques are an extension of ordinary least squares regression in which we iteratively update the minerals incorporated as features in the model with the goal of minimizing error. With FSR, we begin with an empty set and increase the model size based on the p-value for updating the model. Alternatively with BSR, the entire spectral library is included in the set of model features and we decrease the model size. We conducted OLS on the model set of minerals deemed most significant after FSR and BSR to estimate the fractional abundances.

\subsection{Bayesian Model Averaging}
BMA is a technique in which we ensambled a multitude of NNLS regression models. In this study two techniques for BMA were explored -- a basic model and quadratic (BMA-Q) model. In our implementation of BMA-Q, we experimented with second order terms to determine if higher accuracy can be achieved through modeling the interaction between two materials. We coded both BMA and BMA-Q techniques independently rather than using a python package.

\label{results}
\begin{table*}[h!] 
\begin{center}    \includegraphics[width=\textwidth]{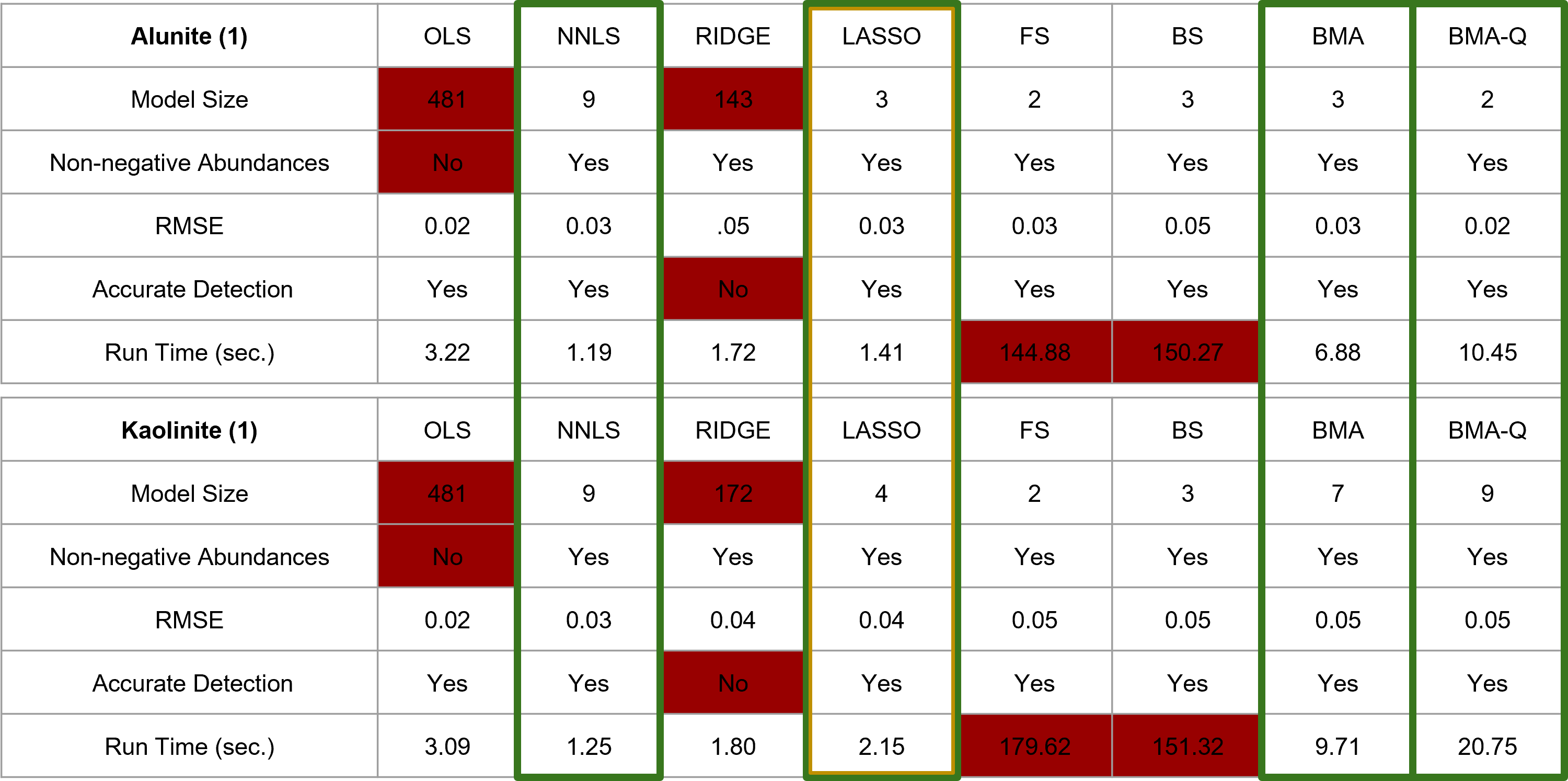}
\caption{\label{compresults} This table displays the comprehensive results of the unmixing techniques.
}
\end{center}
\end{table*}

\begin{figure*}[h!] 
\centering
\begin{subfigure}{.43\textwidth}
\includegraphics[width=\textwidth]{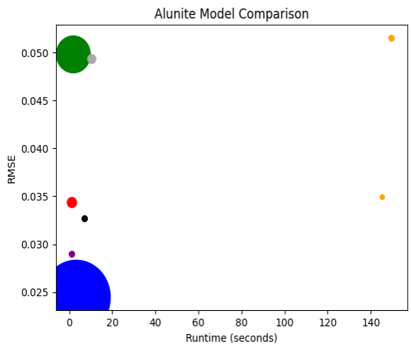}
\caption{}
\label{fig:alunite}
\end{subfigure}
\hspace{.2cm}
\begin{subfigure}{.53\textwidth}
\includegraphics[width=\textwidth]{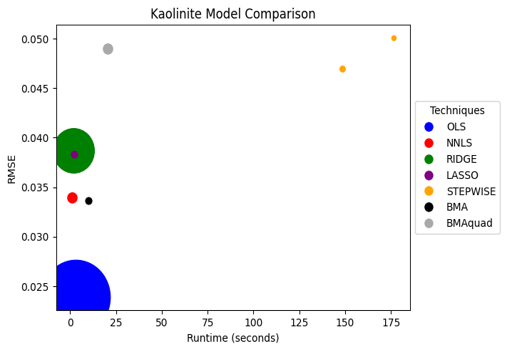}
\caption{}
\label{fig:kaolinite}
\end{subfigure}
\caption[example] { \label{fig:mixturemodelcomare} 
The figures show a comparison of the results based on RMSE, runtime and model size . Figure~\ref{fig:alunite} displays the results for alunite. Figure~\ref{fig:kaolinite} displays the results for kaolinite.}
\end{figure*}

\section{Results}

Table \ref{compresults} and figure \ref{fig:mixturemodelcomare} display the comprehensive results and comparison for alunite hill 1 and kaolinite region 1. In table \ref{compresults}, the performance of each technique is displayed in terms of model size, non-negative abundance incorporation, error, accurate mineral detection and computation time. Figure \ref{fig:mixturemodelcomare} displays the model error on the y-axis, model run time on the x-axis and the model size by the dot diameter. We stored model sizes values, inferred features and abundances for each pixel observation. Results were computed and displayed based on the most common model size for each of the ROIs. The most common model size was used to determine the threshold for including minerals in the regional model. Based on the model size threshold, we selected minerals most commonly incorporated in the individual pixel models. The average abundance of the minerals included the regional model was then calculated to determine the regional y-infer. We computed the RMSE between the regional y-infer and each y-infer at the individual pixel level. Table \ref{compresults} and figure \ref{fig:mixturemodelcomare} show the lowest calculated RMSE of the all the pixels in the ROI.

OLS and ridge regression performed the worst in terms of unmixing the range of pixels for both alunite hill 1 and kaolinite region 1. Though OLS had a relatively fast run time, low error (RMSE = 0.02 for both ROIs) and positive detection of the target mineral, this technique incorporated negative abundance values in the models and included all the spectra from the library in the model. Ridge regression performed similarly to OLS with large model sizes (showing evidence of overfitting within y-infer). Additionally with ridge regression, we failed to detect the target mineral at the .1 abundance threshold for any of the individual pixel models.

The iterative approaches achieved better results with smaller model sizes, exclusive incorporation of non-negative abundances and detection of the target mineral. FSR had a regional model size of two minerals and BSR had a model size of three -- substantially smaller than OLS and Ridge regression. Despite these performance improvements, the computation time for these iterative approaches increased from seconds to over two (alunite hill 1) or three (Kaolinite region 1) minutes.  

NNLS performed well and achieved the lowest run time (1.19 seconds for alunite hill 1). This technique soley incorporated non-negative abundances, computed low error, and detected the target minerals. The model sizing for the NNLS regional models -- 9 minerals -- were smaller than OLS and ridge regression models, but larger than models for all the other techniques.

The BMA approaches had similar results to NNLS but incorporated fewer minerals in their models. Though the BMA approaches achieved useful results, the computation time for BMA and BMA-Q increased, ranging from 6.88 seconds (alunite) to 20.75 seconds (kaolinite). BMA-Q achieved slightly better results for the alunite ROI but worse in terms of model size for the kaolinite ROI.

Lasso regression performed the best across all the techniques. With this technique, we quickly detected  (1-2 seconds) the target minerals with small model sizes, low RMSE and non-negative associated abundances.

\section{Discussion}
\label{discussion} 

This study aimed to unmix complex mixtures using a variety of OLS-based techniques. OLS is a common interpretation of the liner mixture model but can result in impractical solutions or a non-invertible matrix from large spectral libraries. A common challenge associated with intimate mixtures is the high within class and external class variability from the image and thus a need for a large spectral library. We presented methodologies that modify the OLS technique to ensure positive abundance values, penalize model size and or ensamble multiple methods.

Ridge regression was the only technique that performed poorly besides OLS. The L2 norm penalty assigned to the abundances causes the values to shrink and become smaller, but not necessarily zero. So even with a positively bounded solution space, the fractional abundances resulted in very small decimals for all materials in the library to include the target mineral. This regularization attribute caused model sizes to remain large and substantially overfit the observed spectra.

FSR and BSR, preformed better with detecting the target minerals. Additionally, the algorithm design for both techniques ensure non-negative abundances. These techniques also showed to have small model sizes. However, the iterative nature of FSR and BSR affected computation time. Unmixing one pixel provides fairly quick results but because we calculate the p-values of 481 models (for each pixel in in the ROI), the computation time is compounded.

We achieved fairly good results with NNLS and the BMA techniques. The BMA techniques achieved better results for the aluntie ROI in terms of model sizing. NNLS had a faster run time than the BMA techniques and lower error with regard to the kaolinite ROI. The longer computation time experienced with the BMA techniques resulted from ensambling thousands of NNLS models per pixel.

In the context of unmixing intimate mixtures such as minerals, we recommend Lasso regression. We achieved fast and accurate results using this techniques. The L1 norm penalty shrinks abundance values toward zero resulting in successful sparse regression. Thus, lasso regression performed well despite the large spectral library needed for unmixing complex mixtures. We were able to detect the primary mineral with small model sizes for each pixel. Moreover, the computation time remained low regardless of the number of pixels in the ROI. In future research, we recommend ensambling lasso regression model with the BMA techniques -- instead of NNLS -- to evaluate further performance increases.


\bibliography{refs.bib} 
\bibliographystyle{IEEEbib} 

\appendix 
\section{Referenced code}
All the referenced, additional and supporting code will be stored within the corresponding GitHub repository (\url{https://github.com/bakerjf1993/Hyperspectral-Unmixing}).

\end{document}